\documentclass[twocolumn]{aastex631}

\usepackage{natbib}

\usepackage[T1]{fontenc}
\usepackage{longtable}
\usepackage{xhfill}

\setcitestyle{authoryear,open={(},close={)}} 

\begin{document}

\def\msun{M$_\odot$}
\def\iso#1{$^{#1}$}

\title{An updated picture of pre-solar history from short-lived radioactive isotopes and inferences on the birth of the Sun} 


\author{B. So\'os}

\affiliation{Konkoly Observatory, HUN-REN CSFK, Konkoly Thege Mikl\'os  \'ut 15-17, Budapest, H-1121, Hungary}

\affiliation{MTA Centre of Excellence, Konkoly Thege Mikl\'os  \'ut 15-17, Budapest, H-1121, Hungary}

\author{T. C. L. Trueman}

\affiliation{University of Bayreuth, BGI, Universitätsstraße 30, 95447 Bayreuth, Germany}

\affiliation{Konkoly Observatory, HUN-REN CSFK, Konkoly Thege Mikl\'os  \'ut 15-17, Budapest, H-1121, Hungary}

\affiliation{MTA Centre of Excellence, Konkoly Thege Mikl\'os  \'ut 15-17, Budapest, H-1121, Hungary}

\author{A. Yag\"ue L\'opez}

\affiliation{Los Alamos National Laboratory, Los Alamos, NM 87545, USA}

\author{L. Roberti}

\affiliation{Istituto Nazionale di Fisica Nucleare - Laboratori Nazionali del Sud, Via Santa Sofia 62, Catania, I-95123, Italy}

\affiliation{Konkoly Observatory, HUN-REN CSFK, Konkoly Thege Mikl\'os  \'ut 15-17, Budapest, H-1121, Hungary}

\affiliation{MTA Centre of Excellence, Konkoly Thege Mikl\'os  \'ut 15-17, Budapest, H-1121, Hungary}

\author{M. Lugaro}

\affiliation{Konkoly Observatory, HUN-REN CSFK, Konkoly Thege Mikl\'os  \'ut 15-17, Budapest, H-1121, Hungary}

\affiliation{MTA Centre of Excellence, Konkoly Thege Mikl\'os  \'ut 15-17, Budapest, H-1121, Hungary}

\affiliation{Institute of Physics and Astronomy, ELTE Eötvös Loránd University, Budapest, Hungary}

\affiliation{School of Physics and Astronomy, Monash University, VIC 3800, Australia}


\begin{abstract}

We examine the origin of the short-lived radionuclides (SLRs, defined as having half-lives between 0.1 and 100 Ma) present in the early Solar System (ESS) by investigating how predictions of their abundances in the interstellar medium (ISM) from steady-state equilibrium relate to their ESS values. For this, we take into account the non-negligible time $t_{\mathrm{iso}}$ elapsed between the isolation of the pre-solar molecular cloud and the formation of the ESS, during which the SLRs decayed freely. We also consider the alternative scenario in which the pre-solar molecular cloud remained partially mixed with the ISM, with a mixing timescale $t_{\mathrm{mix}}$. We find that the ESS abundances of \iso{107}Pd and \iso{182}Hf produced by \textit{slow} neutron captures (\textit{s}-process), and of \iso{53}Mn and \iso{60}Fe produced by explosive nucleosynthesis, can be consistently explained within these scenarios. Their required $t_{\mathrm{iso}}$ is 9-12 Ma, and their required $t_{\mathrm{mix}}$ is 11-14 Ma (with one potential exception of $t_{\mathrm{mix}}$ = 38 Ma), depending on galactic uncertainties, such as the galactic star formation history and efficiency and the star-to-gas mass ratio. Another \textit{s}-process SLR, \iso{205}Pb has a more uncertain ESS value, and falls within only some of these time values. The same applies to the SLRs produced by the $p$-process (\iso{92}Nb and \iso{146}Sm), depending on the latter's half-life. In agreement with previous studies, we find that the ESS abundances of the \textit{rapid} neutron-capture isotopes (\iso{129}I, \iso{244}Pu, and \iso{247}Cm) and of the most short-lived radionuclides (\iso{26}Al, \iso{36}Cl and \iso{41}Ca) cannot be explained by assuming steady-state equilibrium in the ISM.

\end{abstract}


\section{Introduction}
\label{sec:intro}

The presence of several short-lived radionuclides (SLRs) in the early Solar System (ESS) is testified by meteoritic analysis and provides crucial insights into the formation and early evolution of the Solar System \citep{dauphas2011, davis2022}. The SLRs are characterised by mean lifetimes ($\tau$ = T$_{1/2}$/ln(2), where T$_{1/2}$ is the lifetime) from a few hundred thousand to a hundred million years, and are synthesised by various nucleosynthetic processes occurring in diverse astrophysical environments (see also Table~\ref{tab:origins}). Explosive nucleosynthesis in different types of supernovae (SNe) produces the SLRs \iso{53}Mn and \iso{60}Fe \citep{trueman2025}. Various flavours of the \textit{p}-process (e.g., photo-disintegration) in Type Ia SNe and core-collapse SNe (CCSNe) play the dominant role in the production of \iso{92}Nb and \iso{146}Sm \citep{travaglio2014, travaglio2018}. The \textit{slow} neutron-capture process (the \textit{s}-process) in asymptotic giant branch (AGB) stars plays a dominant role in the production of other SLRs, like \iso{107}Pd, \iso{182}Hf, and \iso{205}Pb \citep{Trueman_2022,leckenby24}, while other heavy SLRs (\iso{129}I, \iso{255}Pu, and \iso{247}Cm) are dominantly made by the \textit{rapid} neutron-capture process, or the \textit{r}-process \citep{Cote2019a, Lugaro2022}. The most short-lived SLRs (\iso{26}Al, \iso{36}Cl, and \iso{41}Ca) are ejected by both the winds of Wolf-Rayet (WR) stars \citep{brinkman2019, brinkman2021, brinkman2023} and by CCSN explosions \citep{limongi2006, Lawson2022}, with the latter being the dominant contributor overall in the Galaxy. Irradiation by solar energetic particles in the ESS could also have produced these three SLRs, and a few others (such as \iso{10}Be, which is not typically made in stars, and \iso{53}Mn). However, irradiation probably played a limited role, except for possible late irradiation in case of \iso{36}Cl. This is because it predicts the abundance of \iso{10}Be to be much higher than that of any other SLRs, and all the SLR abundances to be heterogeneous, both contrary to the observations \citep[see discussion and references in, e.g.][]{desch2023}.

When production and decay balance out, the abundances of SLRs achieve a steady-state equilibrium in the interstellar medium (ISM). One way to check for consistency among constraints from all the different SLRs is to plot the predicted and observed radioactive-to-stable abundance ratios in the ESS divided by their stellar production ratios, against the respective mean life of the SLRs. For examples, see Fig. 5 of \citet{Huss2009}, its updated version in Fig. 10 of \citet{lugaro2018radioactive}, Fig. 6 of \citet{desch2023}, and Fig.~\ref{fig:fig1} of this paper. This representation allows us to check how many observed data points are explained self-consistently by any given scenario. This type of representation has been widely used, as it is very useful in showing the overall picture and checking for consistency among SLRs. However, it has a few disadvantages, which hinder accurate interpretation. For example:

\begin{enumerate}
\item[(1)] Uncertainties related to the evolution of the Milky Way Galaxy are usually not considered. The galactic star formation history and efficiency, star-to-gas mass ratio, and galactic outflows, for example, influence the abundances of the stable and long-lived reference isotopes \citep{Cote2019a}.
\item[(2)] The steady-state formula is used in predicting the abundances of SLRs in the ISM (as will be described in detail in Sec.~\ref{sec:methodsteady}). However, due to statistical variations, uncertainties are expected for the ISM abundances \citep{Cote2019b}. Moreover, below a certain threshold in the ratio of mean life to recurrence time (the time between two injections of the SLRs), the steady-state simply cannot be applied since the SLR may decay entirely before the next production event.
For some isotopes, a steady-state solution does not even exist because the events that produce them are too rare, as demonstrated by \citet{Cote_2021} for the $r$-process SLR \iso{129}I and \iso{247}Cm. 
\item[(3)] There are two main problems with the stellar production ratios:
\begin{itemize}
    \item[(a)] 
These ratios must be chosen as a constant, representative value. However, several distinct astrophysical sites can synthesise the same isotopes with varying contributions as well as recurrences (see Table~\ref{tab:origins}). Furthermore, even enrichment events of the same type can present a range of production ratios, rather than a single value. 
\item[(b)] Moreover, the used values for the stellar production ratios often do not reflect the most recent available literature. For example, values from the analytical description of galactic chemical evolution (GCE) of \citet{Huss2009}, based on stellar yields from \citet{woosley1995}, are still used, together with $r$-process isotopic ratios, e.g., from \citet{goriely2001}. Moreover, the separation of the $s$- and the $r$-processes \citep{Lugaro_2014} is not taken into account.
\end{itemize}
\end{enumerate}

To address these problems, we provide here an updated version of the figure of the radioactive-to-stable ratios, divided by their stellar production ratios as a function of $\tau$, based on the most recent results regarding the stellar production of the SLRs \citep{Cote2019a, Lawson2022, leckenby24} and their GCE \citep{travaglio2014, Trueman_2022, trueman2025}. We provide all the updated quantities, by which we have our figures, and the Python code to build the figure, to facilitate communication between cosmochemistry and nuclear astrophysics and to make it easier to interpret future meteoritic measurements. In Sec.~\ref{sec:methodsteady} we describe the theory of the steady-state solution, outline its limitations, and explain how we handle Problem (1).Section~\ref{sec:timescales} addresses Problem (2), then describes the timescales associated with the ESS that control the relationship between the isotopic ratios in the ISM and in the ESS. In Sec.~\ref{sec:methodinput}, we present our stellar input and address Problem (3). In Sec.~\ref{sec:resanddis}, we present and discuss the updated figures. In Sec.~\ref{conclusion}, we summarise our key results and future work. Note that a recent analysis, using full probability distribution functions rather than error bars, complementary to the results shown here, can be found for selected isotopes in \citet{leckenby24} and in \citet{trueman2025}.


\section{Methodology}
\label{sec:method}

In this section, we describe in detail the methodology used to construct all our figures, from the equations (Sec.~\ref{sec:methodsteady}) to all the necessary inputs (Sec.~\ref{sec:methodinput}). The open-source (CC-BY-4.0 licence) Python code we have developed to build the figures is available on GitHub: \dataset[data for Soós et al. (2026)]{https://github.com/bsos212/Soos-et-al.-2025-dataset}.

\subsection{The steady-state scenario and its limitations}\label{sec:methodsteady}

Instead, if the mean lifetime of the SLR is much shorter than the average time interval between their injections ($\tau/\delta \lesssim 0.3$), then it is statistically likely that the SLR decays significantly before a new enrichment event \citep{Cote2019b}. Therefore, there is a high probability that only one enrichment event left its signature in the ESS\footnote{Note that cases where $0.3<\tau/\delta<2$ require a small number of events and a dedicated statistical approach, which is currently lacking.} \citep[][]{Cote_2021}.
 

The absolute abundance of SLRs is not directly measurable in the ESS, but their ratios relative to a stable or long-lived isotope of reference can be obtained from the meteoritic data. Therefore, it is necessary to calculate these isotopic ratios in the ISM. As discussed in detail in \citet{clayton1988}, the isotopic ratios are time-dependent because while the abundance of the unstable isotope remains close to its steady-state equilibrium value, the abundance of the stable isotope increases throughout the lifetime of the Galaxy. The time evolution of the abundance of the radioactive isotope ($Z_R$) is given by the following formula:
\begin{equation}
    \frac{dZ_R}{dt}= -\frac{Z_R}{\tau}+ \frac{dP_R}{dt},
    \label{eq:eq1}
\end{equation}

\noindent where the first term on the right side represents radioactive decay, and the second represents the supply from stellar sites described by a continuous production rate. When the radioactive isotope reaches equilibrium, then its abundance no longer changes, which means that the left side of the equation becomes zero ($dZ_R/dt=0$), and the production and decay rates are in balance:

\begin{equation}
    \frac{Z_R}{\tau} = \frac{dP_R}{dt}.
    \label{eq:eq2}
\end{equation}

The stable reference isotope instead does not decay; therefore, the equation for its ISM abundance (assuming that $dP_{\mathrm{ref}}/dt$ is constant), $Z_{\mathrm{ref}}$ is the following:

\begin{equation}
    Z_{\mathrm{ref}} = \frac{dP_{\mathrm{ref}}}{dt}T_{\mathrm{gal}},
    \label{eq:eq3}
\end{equation}

\noindent where $dP_{\mathrm{ref}}/{dt}$ is the continuous production rate and $T_{\mathrm{gal}}$ is the age of the Galaxy at the time of the formation of the Solar System, here is taken to be $T_{\mathrm{gal}}$= 8.5 Gyr. If we divide Eq. (1) by Eq. (3), and consider that 
$(dP_R/dt)/(dP_{\mathrm{ref}}/dt)$ can be described by the constant ratio of the stellar yield from the given source ($P_R/P_{\mathrm{ref}}$), we obtain:

\begin{equation}
   \left(\frac{Z_R}{Z_{\mathrm{ref}}}\right)_{\mathrm{ISM}} = \frac{P_R}{P_{\mathrm{ref}}} \frac{\tau}{T_{\mathrm{gal}}},
    \label{eq:eq4}
\end{equation}

\noindent which provides a simple way to represent the ISM equilibrium value of the isotopic ratio $(Z_R/Z_{\mathrm{ref}})_{\mathrm{ISM}}$ as a function of the ratio of the production rates, $T_{\mathrm{gal}}$, and $\tau$.

In this configuration, however, the steady-state formula does not take into account the effect of the galactic star formation history and efficiency, star-to-gas mass ratio, and galactic outflows, all of which can alter the abundance of the stable isotopes \citep{clayton1988, Huss2009}. Full galactic chemical evolution (GCE) models can estimate the uncertainties arising from these factors \citep{Cote2019a}. These effects can be incorporated into the formula via a multiplicative factor $K$, such that:

\begin{equation}
    \left(\frac{Z_R}{Z_{\mathrm{ref}}}\right)_{\mathrm{ISM}} = K\frac{P_R}{P_{\mathrm{ref}}} \frac{\tau}{T_{\mathrm{gal}}}.
    \label{eq:eqSS}
\end{equation}

\noindent Hereafter, we will refer to Eq. \ref{eq:eqSS} as the steady-state formula. \citet{Cote2019a} found the value of $K$ (for ratios that involve a stable isotope) to be 2.3$^{+3.4}_{-0.7}$, where the error bar of $K$ represents the uncertainties in the galactic effects mentioned above. 

There are certain SLRs (\iso{247}Cm and \iso{244}Pu) whose elements lack stable isotopes; therefore, their abundances are measured relative to a long-lived isotope (\iso{235}U and \iso{238}U, respectively) instead of a stable reference isotope. This alters both the steady-state formula and the value of $K$. In fact, the steady-state equation (Eq. \ref{eq:eqSS}) presupposes the full stability of the reference isotope (in Eq. \ref{eq:eq3}), while in these cases, the reference isotopes do not simply accumulate, but also evolve toward a steady-state. Hence, in the ISM, the behaviour of their abundance ($Z_{\mathrm{ref}}$) does not follow Eq. \ref{eq:eq3}, but can be described by applying Eq. \ref{eq:eq2}, modulated via the appropriate value of $K$ calculated using full GCE models, and measuring how their abundances deviate from the steady-state and approximate the behaviour of a stable isotope \citep{Cote2019a}. This leads to 

\begin{equation}
    \left(\frac{Z_{R}}{Z_{\mathrm{ref}}}\right)_ {\mathrm{ISM}}=K\frac{P_{R}}{P_{\mathrm{ref}}} \frac{\tau_{R}}{\tau_{\mathrm{ref}}},
\end{equation}

\noindent where $\tau_{\mathrm{ref}}$ represents the mean life of the reference isotopes and $K$ here takes on different values, which are computed using GCE models for each specific ratio (see Table \ref{table:K}). This \textit{K}-factor is a way to address Problem (1), presented in the Introduction.

Rearranging the steady-state equation by moving the $P_R/P_{\mathrm{ref}}$ ratio from the right to the left side, we obtain the steady-state ratios as a function of $K$, $\tau$, and $T_{\mathrm{gal}}$ (8.5 Gyr), plotted as the red line in Fig.~\ref{fig:fig1}. As already mentioned, the steady-state equation and the value of $K$ are different for \iso{244}Pu and \iso{247}Cm; in these cases, the steady-state value is higher than the plotted line, and is instead shown by red triangles.

\subsection{Timescales of the ESS}\label{sec:timescales}

Another important consideration is that, in reality, the steady-state values are not constant in time or space, because the injection of the material into each parcel of the ISM is not continuous but occurs in discrete stochastic stellar events. The SLRs can freely decay between these injections, causing oscillations around the equilibrium values. \citet{Cote2019b} estimated these deviations using a statistical Monte Carlo approach, and their results are represented by the red shaded area around the red line in Fig. \ref{fig:fig1}. All in all, this method addresses Problem (2) of the Introduction.

Here, we specifically adopt the uncertainties derived assuming $\tau/\delta$=3.16 and its corresponding values from Table 3 of \citet{Cote2019b}, which are the largest reported in that table under the assumption of steady-state equilibrium\footnote{Note that if we extrapolate the trend from $\tau/\delta$=10 and 3.16 to 2 (the actual lowest limit for steady state), the uncertainty changes very little, and also note that we used the “box\_50Myr\_10Gyr” case.}. Even though the value of $\delta$ for which $\tau/\delta$ is 3.16 depends on $\tau$, the uncertainty does not change with it, as it depends only very weakly on the $\delta$ itself \citep[see Table 3 of][]{Cote2019b}. For example, for \iso{53}Mn and \iso{60}Fe, the corresponding $\delta$ is around 1-2 Ma, while for \iso{107}Pd and \iso{182}Hf, the $\delta$ corresponding to $\tau/\delta$=2 is around 3-4 Ma, but the uncertainty does not change with $\delta$. It should be kept in mind that the true $\delta$ may be larger than these values \citep[][found values around 10 Ma for CCSNe]{Wehmeyer_2023}, which would not satisfy the steady-state assumption. However, these authors did not report a $\delta$ for AGB stars, nor did they include diffusion in their models, which would decrease $\delta$.



The next step is to consider the isolation time ($t_{\mathrm{iso}}$) that elapsed between the formation of the giant molecular cloud (GMC) and the formation of the first solids in the ESS. During this time interval, the mixing of materials between the hotter ISM and the cooler GMC may be inefficient, as the mixing timescale is probably slow \citep[$\sim$100 Ma,][]{deAvillez_2002}. If mixing is neglected, the radioactive nuclei decay without resupply from new stellar injections. Consequently, the ESS did not directly inherit the ISM steady-state abundances, and SLRs decayed exponentially during the time interval:

\begin{equation}
    \left(\frac{Z_R}{Z_{\mathrm{ref}}}\right)_{\mathrm{ESS}}= \left(\frac{Z_R}{Z_{\mathrm{ref}}}\right)_{\mathrm{ISM}} \times e^{-t_{\mathrm{iso}}/\tau},
    \label{eq:eq7}
\end{equation}

\noindent where $t_{\mathrm{iso}}$ corresponds to the free-decay time. The blue lines in Fig.~\ref{fig:fig1} represent this scenario with decay times of 10, 20, and 30 Ma.
In the case of SLRs with an unstable reference isotope (\iso{244}Pu and \iso{247}Cm), the decay results are instead indicated by blue triangles. Note that in this equation, the unstable reference isotopes do not decay during the isolation time; however, given their very long half-lives, their decay would increase the ratio by less than 2\%.
 

In the alternative model proposed by \citet{Clayton1983}, mixing is not neglected. In this model, the ISM is composed of three phases: molecular clouds, large HI clouds, and small HI clouds, which communicate and exchange material on a timescale of 10 to 100 Ma. In this scenario, the $Z_R/Z_{\mathrm{ref}}$ ratio does not decrease exponentially, but according to the following quadratic factor\footnote{For further discussion see: Sec. 2 of \citet{Huss2009}, the equation provided here is Eq. (4.3) of \citet{Rauscher2013}.}:
\begin{equation}
    \left(\frac{Z_R}{Z_{\mathrm{ref}}}\right)_{\mathrm{ESS}} = \left(\frac{Z_R}{Z_{\mathrm{ref}}}\right)_{\mathrm{ISM}}  
    \left(1+1.5 \frac{t_{\mathrm{mix}}}{\tau}+0.4(\frac{t_{\mathrm{mix}}}{\tau})^2\right),
    \label{eq:eq8}
\end{equation}

\noindent where $t_{\mathrm{mix}}$ is the mixing timescale after which all the material in a molecular cloud is exchanged with the surrounding region. The blue lines in Fig.~\ref{fig:fig3} represent this scenario with mixing times of 10, 40 and 100 Ma. All the lines representing the free decay in Fig.~\ref{fig:fig1} and the mixing scenario in Fig.~\ref{fig:fig3} inherit the uncertainty of the steady-state line, as indicated by the blue shaded regions surrounding them.

\begin{table}[]
    \centering
    \begin{tabular}{c c c c c}
      Reference isotope & $\tau_{\mathrm{ref}}$ (Ma) & $K_{\mathrm{min}}$& $K_{\mathrm{best}}$& $K_{\mathrm{max}}$\\ \hline
      \iso{235}U & 1016 & 1.1&1.2&1.8\\
      \iso{238}U & 6447 & 1.5&1.9&4.1\\
      Stable isotopes & - & 1.6&2.3&5.7\\
    \end{tabular}
    \caption{The best values of $K$ and their allowed minimum and maximum values, as derived from full GCE calculations, and for the different reference isotopes, from \cite{Cote_2021}, and corrected by \cite{Lugaro2022}.}
    \label{table:K}
\end{table}


\begin{table*}[]
\begin{center}    
\caption{Stellar sites of origin of the SLRs considered here and their stable reference isotopes. Minor components ($\lesssim $10\%) are indicated in brackets.}
    \centering
    \begin{tabular}{c c c c}
    SLR& Origin&Ref.&Origin\\ \hline
    \multicolumn{4}{c}{Very short-lived SLRs} \\ \hline
    \iso{26}Al  & CCSNe (massive star winds, novae, AGBs) & \iso{27}Al & CCSNe \\
    \iso{36}Cl  & CCSNe (massive star winds, AGBs)        & \iso{35}Cl & CCSNe \\
    \iso{41}Ca   & CCSNe (massive star winds, AGBs)       & \iso{40}Ca & CCSNe \\ \hline
    \multicolumn{4}{c}{Explosive SLRs}\\ \hline
    \iso{53}Mn  & CCSNe, Type Ia SNe  & \iso{55}Mn & CCSNe, Type Ia SNe \\
    \iso{60}Fe  & CCSNe (AGBs, ECSNe$^a$) & \iso{56}Fe & CCSNe, Type Ia SNe \\         \hline
    \multicolumn{4}{c}{$s$-process SLRs$^b$}\\ \hline
    \iso{107}Pd & AGBs, CM$^c$, Rare CCSNe$^d$ & \iso{108}Pd & AGBs, CM, Rare CCSNe \\
    \iso{182}Hf & AGBs, CM, Rare CCSNe & \iso{180}Hf & AGBs, CM, Rare CCSNe \\
    \iso{205}Pb & AGBs                 & \iso{204}Pb & AGBs \\ 
    \hline
    \multicolumn{4}{c}{$p$-process SLRs}\\ \hline
    \iso{92}Nb  & CCSNe, Type Ia SNe & \iso{92}Mo & CCSNe, Type Ia SNe \\
    \iso{146}Sm & CCSNe, Type Ia SNe & \iso{144}Sm & CCSNe, Type Ia SNe \\
    \hline
    \multicolumn{4}{c}{$r$-process SLRs}\\ \hline
    \iso{129}I  & CM$^c$, Rare CCSNe$^d$ & \iso{127}I & CM, Rare CCSNe \\
    \iso{244}Pu & CM, Rare CCSNe & \iso{238}U & CM, Rare CCSNe \\
    \iso{247}Cm & CM, Rare CCSNe & \iso{235}U & CM, Rare CCSNe \\
    \end{tabular}
    \label{tab:origins}
\end{center}
$^a$ Electron-capture supernovae. $^b$The SLRs \iso{107}Pd and \iso{182}Hf are also produced by the $r$-process; however, this process did not contribute to their ESS abundances, as the last $r$-process event occurred 100-200 Ma before the Sun's birth. Nevertheless, $r$-process events need to be taken into account as contributors to the abundance of their stable reference isotopes (see Sec.~\ref{sec:resanddis}). $^c$CM = compact mergers, including neutron star–neutron star and neutron star–black hole mergers. $^d$Rare CCSNe are those that lead, for example, to the formation of collapsars and magnetars.
\end{table*}

\begin{table*}
\caption{For each SLR isotope, we list its reference isotope, its mean life in Ma (with 2$\sigma$ error bar) from \citet{lugaro2018radioactive}, except for \iso{146}Sm, from  \citet{tang25}, its ESS values from \citet{lugaro2018radioactive}, except for \iso{60}Fe/\iso{56}Fe, from \citet{fang2025}, and \iso{205}Pb/\iso{204}Pb from \citet{Baker2010}, and the production ratios used in this work, with their reference source.}
    \begin{center}
\begin{tabular}{c c c c c c}
SLR & Ref.& $\tau$ (Ma) & ESS value & Production ratio (PR) & Reference for PR\\ 
\hline
\multicolumn{5}{c}{Very short-lived SLRs}\\
\hline
\iso{26}Al &\iso{27}Al& 1.035(35) & $(5.23 \pm 0.13)\times10^{-5}$ &$(0.19-1.03)\times10^{-2}$&\citet{Lawson2022}\\
\iso{36}Cl &\iso{35}Cl&0.434(3) & $(2.44\pm0.65)\times10^{-5}$     &$(0.65-1.65)\times10^{-2}$  \\
\iso{41}Ca &\iso{40}Ca& 0.143(2) & $4\times10^{-9}$$^a$    &$(0.58-7.55)\times10^{-3}$ \\
\hline
\multicolumn{5}{c}{Explosive SLRs}\\
\hline
\iso{53}Mn &\iso{55}Mn&5.40(6)  & $(7\pm1)\times10^{-6}$           &See Table~\ref{table:exp}               &\citet{trueman2025}\\
\iso{60}Fe &\iso{56}Fe& 3.78(6)  & (7.71 $\pm$ 0.47) $\times 10^{-9}$ & See Table~\ref{table:exp}      & \citet{trueman2025}\\

\hline
\multicolumn{5}{c}{$s$-process SLRs}\\
\hline
\iso{107}Pd &\iso{108}Pd& 9.4(4)  & (6.6$\pm$0.4)$\times 10^{-5}$ & $K$=1.6: 0.097 & \citet{Trueman_2022} \\
&&&&$K$=2.3: 0.099 \\
&&&&$K$=5.7: 0.117 \\
\iso{182}Hf &\iso{180}Hf& 12.8(1.3) & (1.02$\pm$0.04)$\times 10^{-4}$& $K$=1.6: 0.091 & \citet{Trueman_2022} \\
&&&&$K$=2.3: 0.089 \\
&&&&$K$=5.7: 0.100 \\
\iso{205}Pb &\iso{204}Pb& 25.0(1.0) & (1.0$\pm$0.4)$\times10^{-3}$ & 0.167                          & \citet{leckenby24}\\
\hline
\multicolumn{5}{c}{$p$-process SLRs}\\
\hline
\iso{92}Nb &\iso{92}Mo& 50.1(3.5) & (1.57$\pm$0.09)$\times10^{-5}$ & 1.58$\times10^{-3}$$^b$                      & \citet{travaglio2014}\\
\iso{146}Sm &\iso{144}Sm& 103.5(10.7) & (8.28$\pm$0.44)$\times10^{-3}$& 0.347$^b$                         & \citet{travaglio2014} \\
\hline
\multicolumn{5}{c}{$r$-process SLRs}\\
\hline
\iso{129}I &\iso{127}I&22.6(6) & (1.28$\pm$0.03)$\times10^{-4}$ & 1.28  & \citet{Cote2019a} \\
\iso{244}Pu &\iso{238}U&115(1)  & (7$\pm$1)$\times10^{-3}$ &  0.19 & \citet{Lugaro2022} \\
\iso{247}Cm &\iso{235}U&22.5(7) & (5.6$\pm$0.03)$\times10^{-5}$   & 0.30  &\citet{Cote2019a} \\
\end{tabular}
\label{table:big}
\end{center}
$^a$A higher value of 2$\times 10^{-8}$ was reported by \citet{ku2022}, however, given the difficulty in accurately determining this ratio, we used the value recommended by \cite{davis2022}, which is the most commonly observed in CAIs.
$^b$Calculated by \citet{Lugaro_2016} as the average of Table 1 of \citet{travaglio2014} in the range of metallicity 0.01 to 0.02.
\end{table*}

\begin{table*}[]
\caption{The production ratios of explosive SLRs according to different values of $K$, and two different sets of CCSN models. The values in the last two columns (indicated with an asterisk) represent the production ratios calculated by GCE models where the Type Ia SNe yield of \iso{53}Mn was set to zero.  (see Sec.~\ref{sec:exp_input}.).}
\label{table:exp}
\centering
\begin{tabular}{c c c c c c c}
\hline
The value of $K$ &\iso{60}Fe (R)&\iso{60}Fe (N13)&\iso{53}Mn (R)&\iso{53}Mn (N13)& \iso{53}Mn* (R)&\iso{53}Mn* (N13)\\ \hline
$K$=1.6& 1.445$\times10^{-4}$&0.863$\times10^{-4}$&0.1989&0.1871& 0.0790&0.0585\\
$K$=2.3&1.298$\times10^{-4}$&0.804$\times10^{-4}$&0.1493&0.1445& 0.0644&0.0490\\
$K$=5.7& 1.237$\times10^{-4}$&0.748$\times10^{-4}$&0.1552&0.1718& 0.0579&0.0470\\
\end{tabular}
\end{table*}

\subsection{The stellar inputs}\label{sec:methodinput}

Here, we present the stellar input data used to apply the method described above to generate the figures.  In Table~\ref{tab:origins} we list the 13 SLRs that we consider here, with their reference stable isotopes and indicate for each of them the stellar sites that produce significant amounts of each isotope, enough to be included in the GCE modelling \citep[see also][]{lugaro2018radioactive}. Note that the SLRs may have different sites of origin than their reference isotopes.
In Table~\ref{table:big}, we report all the numbers that we used as input for the methodology described above.

\subsubsection{The very short-lived radionuclides: \texorpdfstring{\iso{26}}{Lg}Al, \texorpdfstring{\iso{36}}{Lg}Cl and \texorpdfstring{\iso{41}}{Lg}Ca}

These three isotopes are all produced in massive stars and are ejected by both the winds (in the case of very massive WR stars or binary systems, a minor GCE component) and the CCSN explosion (the major GCE component). In the wind, \iso{26}Al is produced via \iso{25}Mg(p,$\gamma$)\iso{26}Al during H burning, while, \iso{36}Cl and \iso{41}Ca are produced by neutron captures during He burning \citep[e.g.][]{brinkman2023}. The CCSN component of \iso{26}Al comes from the ejection of material from H burning (if not ejected by the previous phases) and C burning, where it is produced by the combined presence of abundant \iso{25}Mg and protons. In this region, protons are produced directly through carbon fusion via the \iso{12}C(\iso{12}C,p)\iso{23}Na reaction and \iso{26}Al can be destroyed by (n,p) and (n,$\alpha$) neutron capture reactions, with neutrons from the \iso{22}Ne($\alpha$,n)\iso{25}Mg reaction \citep[see][for details]{limongi2006, Lawson2022}. As discussed in detail by \citet{Lawson2022}, the CCSN component of \iso{36}Cl and \iso{41}Ca is a combination of their production during He- and C burning, as well as during explosive nucleosynthesis.

Besides these sites of origin, novae have been considered as a source of \iso{26}Al. While \citet{bennett2013} found a high contribution of around 30\%, later studies found this contribution to be less important, \citet{canete2023} found it to be somewhat less than 20\%, and \citet{laird2023} reported $\sim$10\%. This lower contribution factor is also supported by $\gamma$-ray data, which shows that the abundance of \iso{26}Al is highest on the Galactic plane, where younger massive stars are located \citep{diehl06}.

The ranges of production ratios of these three SLRs used as input here (see Table~\ref{table:big}) are from the 62 models of \citet{Lawson2022}, which correspond to three different initial masses (15, 20, and 25 \msun) of solar metallicity. Solar metallicity is appropriate for the SLRs because the yields produced in low-metallicity stars early in the Galaxy would have decayed. However, for the stable isotopes, this is not accurate and full GCE models would be needed, especially for \iso{27}Al and \iso{35}Cl, which are secondary isotopes, i.e., their production is metallicity dependent, unlike \iso{40}Ca, which is primary. In any case, when we compare our results to the full GCE models of \citet{kaur2019} (see Section~\ref{sec:explain-inputs}), we do not find any major difference. In the models, we also considered a variety of explosion energies (from 3.4 × 10$^{50}$ to 1.8 × 10$^{52}$ erg) and compact remnant masses (from 1.5 to 4.89 \msun). This variability represents a way to account not only for the effect of the initial mass but also for systematic uncertainties in CCSN modelling. 

\subsubsection{The explosive SLRs: \texorpdfstring{\iso{53}}{Lg}Mn and \texorpdfstring{\iso{60}}{Lg}Fe } \label{sec:exp_input}

Explosive nucleosynthesis in different types of supernovae produces the SLRs \iso{53}Mn and \iso{60}Fe. \iso{60}Fe is also produced in the pre-CCSN He- and C burning of massive stars \citep{limongi2006,jones2019, Lawson2022}. In all cases, this production happens when the temperature is high enough for the \iso{22}Ne($\alpha$,n)\iso{25}Mg reaction to be activated and release a high-density neutron flux. In these conditions, the unstable \iso{59}Fe (T$_{1/2}$=44.5 days) isotope becomes a branching point, As a result, the production ratio of \iso{60}Fe to \iso{56}Fe is controlled by the ratio of the neutron-capture rates and the $\beta$-decay of \iso{59}Fe as a function of temperature \citep{jones2019}. 

Other proposed sites of neutron-rich nucleosynthesis leading to the production of \iso{60}Fe include electron-capture SNe \citep{wanajo2013} and carbon deflagration Type Ia SNe \citep{woosley1997}. However, these objects are theoretical and rare, so they are not considered here as the main sources of \iso{60}Fe in the Galactic medium. 

The other explosive SLR is \iso{53}Mn, which, unlike \iso{60}Fe, is produced by both CCSNe and Type Ia SNe, with the latter playing a significant role in Mn production \citep{seitenzahl2013_near_CHa, kobayashi2020,gronow2021}. In CCSNe, the majority of \iso{53}Mn and the stable \iso{55}Mn are produced during the explosion by nuclear statistical equilibrium and by explosive O-, Ne-, and/or partial Si burning, depending on the mass of the progenitor star. The amount of ejected \iso{53}Mn is highly dependent on the remnant mass, which controls how much of the inner layers are ejected \citep{Lawson2022}. In Type Ia SNe, both isotopes \iso{53}Mn and \iso{55}Mn are also produced by nuclear statistical equilibrium, with near-Chandrasekhar-mass objects typically resulting in higher Mn abundances and higher \iso{53}Mn/\iso{55}Mn ratios.

Since the two Mn isotopes have two stellar sources, their physical production ratio cannot be represented by one production ratio, as needed for Eq.~\ref{eq:eqSS}. Furthermore, Eq.~\ref{eq:eqSS} is not accurate for \iso{60}Fe either, because its reference stable isotope \iso{56}Fe is also produced by both CCSNe and Type Ia SNe. Therefore, a different approach is needed here. We derived their "effective" production ratio from full GCE models, including both CCSNe and Type Ia SNe \citep{trueman2025}. The GCE models were calculated based on the same framework as \citet{Cote2019a}; therefore, they provide results for the three GCE setups corresponding to the three values of $K$. The GCE models predict the \iso{53}Mn/\iso{55}Mn and the \iso{60}Fe/\iso{56}Fe ratios in the ISM at the Galactic time corresponding to the formation of the Sun, i.e., when the metallicity reaches the solar metallicity. GCE models do not account for the isolation time, given that it is very short relative to the Galactic timescale. Assuming these ISM values in Eq.~\ref{eq:eqSS}, we derive the "effective" production ratio which reproduces the results of the GCE models. This means that we are effectively plotting the results of the full GCE models.

The values of the "effective" production ratios used in this work are listed in Table \ref{table:big}. Due to the many uncertainties in nucleosynthesis in CCSNe, as well as in the origin of Type Ia SNe (in particular, whether they are from the explosion of CO white dwarfs with near-Chandrasekhar or sub-Chandrasekhar mass), many GCE models were calculated by \citet{trueman2025}. For each value of $K$, we used models calculated with different sets of CCSNe: the recommended set of \citet{limongi2018} (R) and the set of \citet{nomoto2013} (N13). For the fraction of sub-Chandrasekhar-mass Type Ia SNe relative to the total number of Type Ia SNe, we used the value of 0.5, but the results would not change substantially when changing this value \citep{trueman2025}.

\citet{trueman2025} also explored GCE models in which the yield of \iso{53}Mn from Type Ia SNe was set to zero. This is to simulate the possibility that the last Type Ia SNe occurred long enough before the formation of the Sun for the vast majority of the ejected \iso{53}Mn to have decayed. This is equivalent to assuming that the contribution of Type Ia SNe to \iso{53}Mn cannot be modelled with the steady-state equation, consistent with the fact that Type Ia SNe are less frequent than CCSNe. In Milky-Way-type galaxies, the rate of CCSNe to Type Ia SNe is observed to be between roughly 4 and 10, including error bars \citep[see, e.g., Table 2 of][]{mannucci2005}, and \citet{bazin09} reports a value of 4.5$\pm$1.4 at red-shift 0.3, i.e., the time of the birth of the Sun. These values correspond to an average $\delta$ of Type Ia SNe at least 3 times longer than that of CCSNe. GCE modelling is necessary even in this case because the stable \iso{55}Mn present in the ISM is also contributed by Type Ia SNe. Also, in this framework, we investigated the impact of using different fractions of sub-Chandrasekhar-mass Type Ia SNe relative to the total number of Type Ia SNe, ranging from 0 to 1, which will be discussed in Sec.~\ref{sec:resanddis}. 

\subsubsection{The \texorpdfstring{$s$}{Lg}-process SLR isotopes: \texorpdfstring{\iso{107}}{Lg}Pd, \texorpdfstring{\iso{182}}{Lg}Hf and \texorpdfstring{\iso{205}}{Lg}Pb}

The ESS abundances of these three SLRs are understood to mainly come from the \textit{slow} neutron-capture process (the $s$-process) that takes place in low- to intermediate-mass stars (1.5 \msun$\lesssim$M$\lesssim$4 \msun) on the AGB \citep[][and references therein]{Trueman_2022,leckenby24}\footnote{\iso{135}Cs also belongs to this group, however, only an upper limit is available for its ESS ratio to \iso{133}Cs. Furthermore, its mean life is 1.9 Ma \citep{MacDonald2016} (note that this is the most recent value, shorter than usually employed), therefore, it is likely that its abundance is not in steady-state. This isotope is discussed in detail in the Supplementary Material of \citet{leckenby24}, and we refer the interested reader to that paper.}. While \iso{205}Pb is an $s$-only isotope (meaning that it is only produced by the $s$-process), the \iso{107}Pd and \iso{182}Hf are also produced by the \textit{rapid} neutron-capture process ($r$-process). For these latter two isotopes, only their $s$-process production is considered here, as the last $r$-process event occurred too long ago to contribute to the ESS inventory (see Sec.~\ref{section:input-r}). 

Stars on the AGB are in their late-stage evolution and have already undergone core H- and He burning. At this stage, an outer H burning shell accumulates He on top of a He-rich intershell, which is itself located above the degenerate C-O core. This accumulated He shell periodically experiences He burning and releases a large amount of energy in a short period of time, in a so-called He-flash or thermal pulse (TP). Between the TPs, the \iso{13}C($\alpha$,n)\iso{16}O and, during the TPs, the \iso{22}Ne($\alpha $,n)\iso{25}Mg reaction channels are activated. Their effectiveness is mainly dependent on the mass and metallicity of the star: in the lower mass range, the former reaction has a greater effect, while in the higher mass range, the latter is more efficient. These reactions produce free neutrons, which other nuclear species can capture, in particular the abundant Fe seeds, leading to the production of the elements up to Pb \citep[see][for a review]{Lugaro2023}. 

In the case of \iso{107}Pd, the stable \iso{106}Pd captures a free neutron to create \iso{107}Pd. The case of \iso{205}Pb is similar (\iso{204}Pb captures a free neutron), except that the strong temperature and density dependence of the electron-capture rate of \iso{205}Pb in to \iso{205}Tl and $\beta$-decay rate of \iso{205}Tl (stable on Earth and produced by a branching point at \iso{204}Tl) make the problem very complex \citep{leckenby24}.
The synthesis of \iso{182}Hf is also complex because (similar to the aforementioned \iso{60}Fe), it occurs through the activation of a branching point at \iso{181}Hf, which has a half-life of 42.3 days. Therefore, the \iso{182}Hf/\iso{180}Hf ratio is controlled by the relative activation of the neutron capture and the $\beta$-decay of this branching point, which in turn depends on the neutron density and the temperature. While the branching point can be activated only at the neutron density produced by the \iso{22}Ne($\alpha$,n)\iso{25}Mg, the \iso{13}C($\alpha$,n)\iso{16}O reaction is also needed for the production of the \iso{180}Hf itself. Therefore, the synthesis of \iso{182}Hf is preconditioned on the activation of both reaction channels, which condition is only present in AGB stars with an initial mass of 3-4 M$_{\odot}$ \citep{Lugaro_2014}.

As in the case of \iso{53}Mn and \iso{60}Fe, also for \iso{107}Pd and \iso{182}Hf, we use "effective" production ratios from the full GCE models of \citet{Trueman_2022}. The ISM ratios predicted by these models already include the $r$-process contribution to the stable reference isotopes \iso{108}Pd and \iso{180}Hf. For \iso{205}Pb instead, we used the stellar production ratio calculated by using the latest nuclear physics input and weighted on a population of AGB stars of solar metallicity \citep{leckenby24}.

As \iso{205}Pb and its stable reference isotope, \iso{204}Pb, are $s$-process-only isotopes \citep[see][]{lugaro2018radioactive, leckenby24}, by comparing their result to those of \iso{107}Pb and \iso{182}Hf, we can test the scenario of the early $r$-process occurrence.

\subsubsection{The \texorpdfstring{$p$}{Lg}-process SLR isotopes: \texorpdfstring{\iso{92}}{Lg}Nb and \texorpdfstring{\iso{146}}{Lg}Sm}

The $p$-process isotopes ($p$-nuclei) are relatively neutron-deficient isotopes that cannot be produced by neutron captures \citep{Thielemann2010, pignatari2016}. Two short-lived $p$-nuclei have been confirmed to have been alive in the ESS: \iso{92}Nb and \iso{146}Sm\footnote{The other two short-lived $p$-nuclei (\iso{97}Tc and \iso{98}Tc) are excluded from this study because only an upper limit is available for their ESS abundances \citep[see][and references therein]{Dauphas2003}, and these upper limits are too high to give meaningful constraints. They also have relatively short mean life ($\sim$4 Ma), and therefore the steady-state scenario may not be accurate.}. The $p$-nuclei can be produced by a variety of processes, and their astrophysical sites of origin are still debated. One of their main production processes is the photo-disintegration ($\gamma$) process, in which heavier isotopes break up into lighter isotopes, including the $p$-nuclei. The $\gamma$-process occurs in CCSNe during explosive O/Ne burning and C-O shell merger events \citep{Roberti_2024}. Charged-particle reactions can also contribute to the production of the $p$-nuclei up to Mo and Ru, for example, during the $\alpha$-rich freeze-out of nuclear statistical equilibrium in CCSNe \citep{woosley1978, woosley1992}, the interaction with matter of the neutrino winds ejected by the nascent neutron stars \citep{ hoffman1996, farouqi2009, arcones2011}, and even during $r$-process events \citep{xiong2024}. The heaviest SLR of the $p$-nuclei, \iso{146}Sm, is more likely to be produced by the $\gamma$-process, whereas the lighter SLR $p$-nuclei are likely to have contributions from charged-particle processes \cite{Lugaro_2016}. The $\gamma$-process can also occur in Type Ia supernovae that originate from accretion onto a C-O white dwarf reaching the Chandrasekhar mass \citep{travaglio2014}. 

\citet{travaglio2014} calculated the GCE of the $p$-process SLR isotopes, assuming that they originated exclusively from Type Ia SNe and found that their ISM abundances at the time of the formation of the Sun are higher than the ESS abundances. \cite{Lugaro_2016} included \iso{53}Mn in the investigation, but did not find a self-consistent solution for the three isotopes (\iso{53}Mn, \iso{92}Nb, \iso{146}Sm), and proposed that the lighter SLR $p$-nuclei (including \iso{92}Nb) should also originate from CCSNe.

As new efforts are still in progress on the analysis of the production of these nuclei in CCSNe \citep{roberti23gamma}, here we decided to assume that the SLR $p$-nuclei originated from Type Ia SNe \citep[i.e., the average of Table 1 of][in the range of metallicity 0.01 to 0.02]{travaglio2014}, and report these production factors as inputs in Table~\ref{table:big}. We note that one of the main obstacles related to the investigation of \iso{146}Sm remains the systematic uncertainty of its half-life. The most recent value of 71.74 $\pm$ 7.39 Myr (at 2$\sigma$) calculated by \cite{tang25} and which we use here, is, in fact, in disagreement with the somewhat higher values reported by two recent experimental studies: 92 $\pm$ 5.2 Myr \citep{chiera2024} and between 82 and 87 Myr \citep{kavner24thesis}. 
 




\subsubsection{The \texorpdfstring{$r$}{Lg}-process SLR isotopes:  \texorpdfstring{\iso{129}}{Lg}I,  \texorpdfstring{\iso{244}}{Lg}Pu, and  \texorpdfstring{\iso{247}}{Lg}Cm}
\label{section:input-r}

These three SLRs are almost exclusively produced by the $r$-process. \iso{244}Pu and \iso{247}Cm are heavier than Pb and Bi, beyond which the $s$-process chain ends, while \iso{129}I cannot be produced by the $s$-process, as the branching point isotope \iso{128}I has a half-life of 25 minutes and, therefore, under $s$-process conditions, it decays before it can capture a neutron.


The production ratio for \iso{129}I/\iso{127}I is derived from the solar $r$-process residual of its stable daughter \iso{129}Xe, while for the other two SLRs, we have chosen typical production values from the reported references. As discussed at length in \citet{Cote_2021}, the $r$-process events are very rare, meaning that it is most likely that for these isotopes $\tau$/$\delta$ is much lower than 2, and these isotopes are not in a steady-state equilibrium. If $\tau$/$\delta$ is lower than 0.3, then they originate from a single last event. Here, we still consider these isotopes to verify whether this is consistent with their location in Figures \ref{fig:fig1} and \ref{fig:fig3}. 

\subsubsection{Further explanations on the choice of the stellar production input}
\label{sec:explain-inputs}

Before moving on to present our results and discuss their implications, we summarise and finalise the explanation of how our work addressed the three problems listed in the Introduction. Problem (1) is addressed by using the $K$ factor from \citet{Cote2019a} to handle galactic uncertainties. Problem (2) is addressed by using the results of \citet{Cote2019b} to quantify the statistical deviation of the isotopic ratios from the equilibrium values in the ISM. Problem (3a) is that the stellar production ratios must be chosen as constant representative values, while they are highly variable. The only way to fully account for this is to run full GCE models and derive "effective" stellar production ratios from them (as explained in Section~\ref{sec:exp_input}). We can follow this approach for \iso{53}Mn, \iso{60}Fe, \iso{107}Pd, and \iso{182}Hf because GCE models available for these isotopes correspond to the three different values of $K$ \citep[i.e., the set-ups used by][]{Cote2019a}. For all the other SLRs, instead, we do not have this information. For the $p$-process SLRs, \citet{travaglio2014} presented results from their GCE model. These are incorporated into our analysis because their results are identical to the values that we obtain when using the minimum value of $K$=1.6\footnote{The values needed to represent the exact results of \citep{travaglio2014}} are K=1.88 and 1.65 for \iso{92}Nb and \iso{146}Sm, respectively.

For the $s$-only \iso{205}Pb, the nuclear-physics input required to calculate its production in AGB stars was measured only recently \citep{leckenby24}. Therefore, there are not enough updated AGB models yet to calculate the full GCE for this isotope. In this case, we resorted to using the best available updated stellar production factor. Therefore, while we could not address problem (3a), we could at least address Problem (3b). Also for \iso{26}Al, \iso{36}Cl, \iso{41}Ca, we could address Problem (3b) by using the latest available models, which also include an investigation of the explosion mechanism \citep{Lawson2022}.

In reference to the latter isotopes, we note that \citet{kaur2019} calculated GCE models for \iso{26}Al, \iso{36}Ca, \iso{41}Cl, \iso{53}Mn, and \iso{60}Fe, of which the last two are in common with \citet{trueman2025}. A full direct comparison is hampered by the different GCE methodology \citep[as in the case of][mentioned above]{travaglio2014}, since \citet{trueman2025} used the three set-ups of \citet{Cote2019a} to create three different Milky Ways corresponding to the three values of $K$, while the four homogenous GCE models of \citet{kaur2019} were created instead using different yield combinations. Still, we found that the results are comparable. When assuming the $K=2.3$ value, the "effective" production ratios from \citet{kaur2019} are between $7.8\times10^{-5}$ and $2.1\times10^{-4}$ for \iso{60}Fe/\iso{56}Fe, a similar range to that derived using \citet{trueman2025} for the same $K$ set-up (see Table~\ref{table:exp}) For \iso{53}Mn, the \citet{kaur2019} "effective" production ratios are between $4.7\times10^{-2}$ and $6.2\times10^{-2}$, i.e., roughly one third of the range derived from the $K=2.3$ model of \citet{trueman2025}. This difference is due to the fact that Type Ia SNe strongly contribute to this isotope, and \citet{trueman2025} updated and tested the impact of new Type Ia SNe yields, while \citet{kaur2019} used the old yields from \citet{iwamoto1999}. For sake of completeness, we also compared the "effective" production ratios from \citet{kaur2019} for the three very short-lived SLRs (\iso{26}Al, \iso{36}Ca, and \iso{41}Cl) to the stellar production ratios from \citet{Lawson2022}, again assuming the $K=2.3$ scenario, and found that the ratios from \citet{kaur2019} are generally a factor of 2-3 higher. This is probably because these authors used different CCSN yields, i.e. from \citet{woosley1995} and \citet{limongi2018}. In any case, this difference does not have any significant effect on our conclusions.


\section{Results and discussion}
\label{sec:resanddis}

\begin{figure*}[ht!]
\centering
\includegraphics[width=\textwidth]{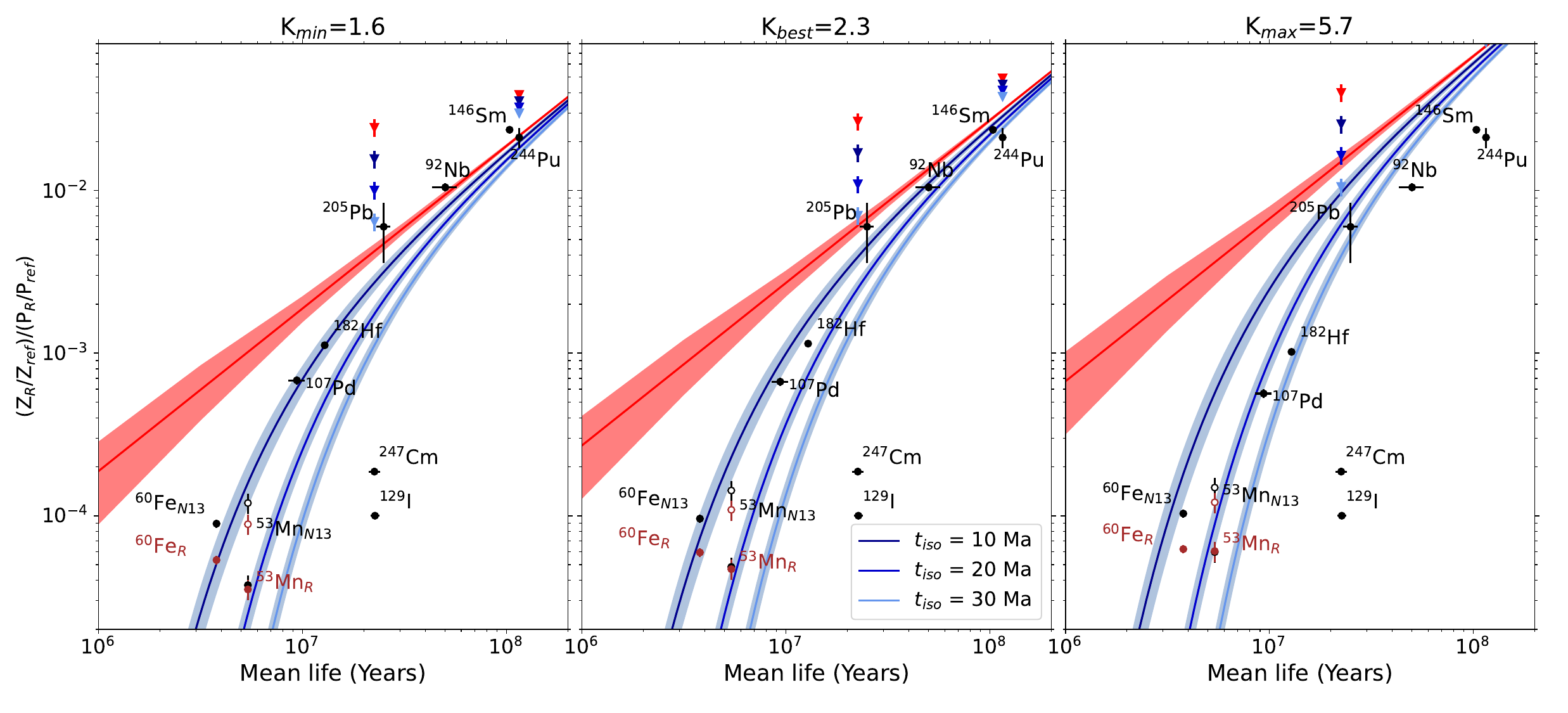}
\caption{Comparison between models and ESS data performed according to the method described in Sec.~\ref{sec:method}. 
The red line represents the abundance ratios of a radioactive isotope relative to stable or long-living isotopes ($Z_R/Z_{\mathrm{ref}}$) divided by the corresponding stellar production ratios ($P_R/P_{\mathrm{ref}}$) as predicted by the steady-state formula (Eq.~\ref{eq:eqSS}) with a Galactic age of 8.5 Ga, as a function of the mean life. The three panels show the values calculated for the three different values of $K$, indicated at the top of each subplot.
The shaded red area represents 1$\sigma$ uncertainties from the Monte Carlo approach of \cite{Cote2019b}, considering $\delta=1$ Ma.
The blue lines represent the results of applying a free-decay time with values indicated in the legend. The downward triangles correspond to the same as the lines above but for \iso{247}Cm and \iso{244}Pu, as these SLRs require a different steady-state equation (see Sec.~\ref{sec:methodsteady}). 
The black circles represent the same ratios as above, except that in this case, $Z_R/Z_{\mathrm{ref}}$ is the ESS value for each SLR as indicated by the corresponding label (for their values see Table~\ref{table:big}). For \iso{53}Mn and \iso{60}Fe, the black and red circles represent, respectively, the GCE results using N13 or R yields. For \iso{53}Mn, the corresponding empty circles represent the GCE models with no \iso{53}Mn ejected by Type Ia SNe. 
\label{fig:fig1}}
\end{figure*}

\begin{figure}[ht!]
\centering
\includegraphics[width=0.45\textwidth]{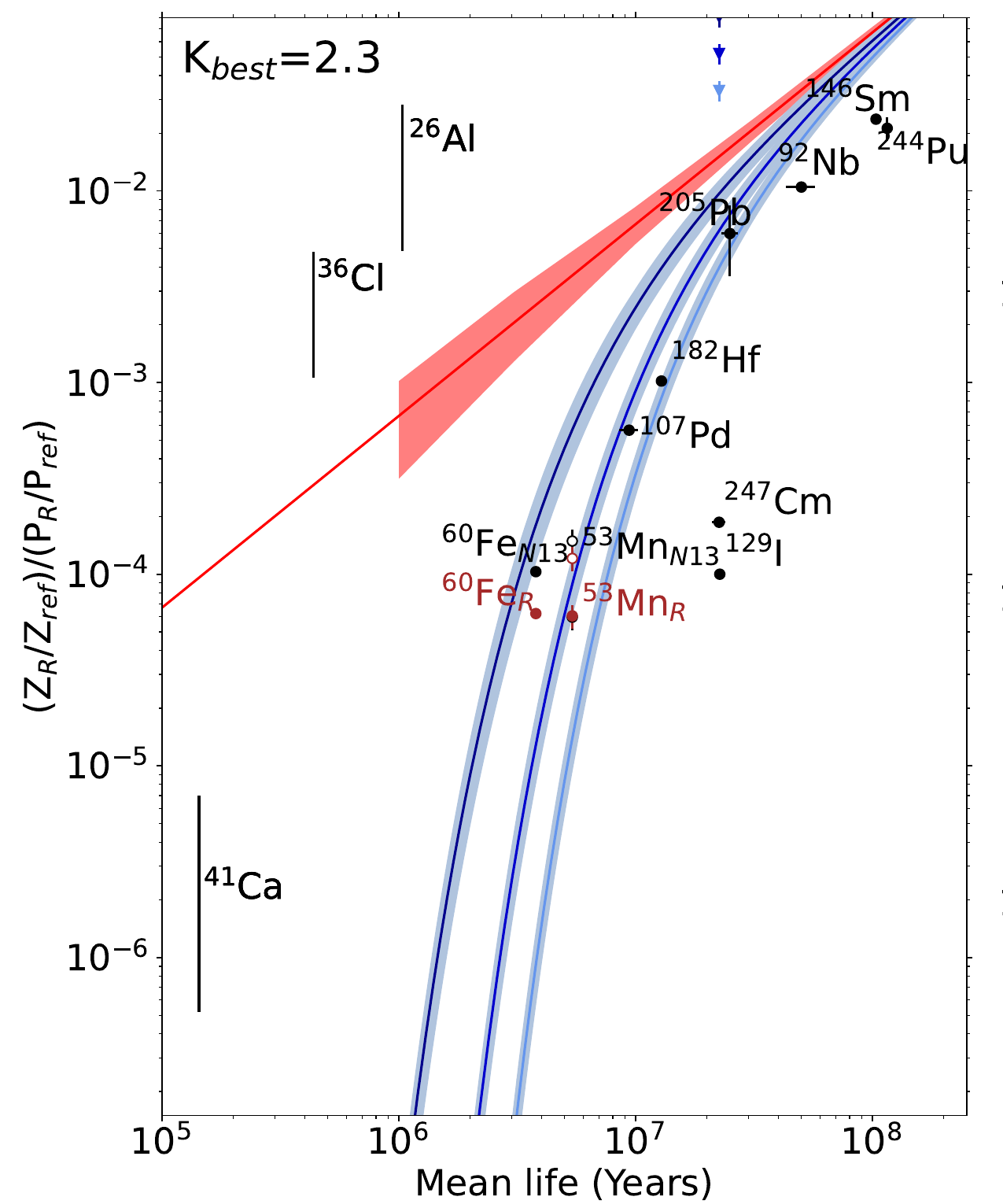}
\caption{Same as Fig.~\ref{fig:fig1} but with $\tau$ extended down to 0.1 Ma to include the shortest-lived SLRs. Vertical lines represent these three SLRs because we used a range of $P_R/P_{ref}$ ratios (see Section~\ref{sec:methodinput}) rather than one value only. The red shaded area is not included below $\tau=1$ Ma because the steady-state approximation is invalid below this value \citep{Cote2019b}.}
\label{fig:fig1zoomout}
\end{figure}

\begin{figure*}[ht!]
\centering
\includegraphics[width=0.99\textwidth]{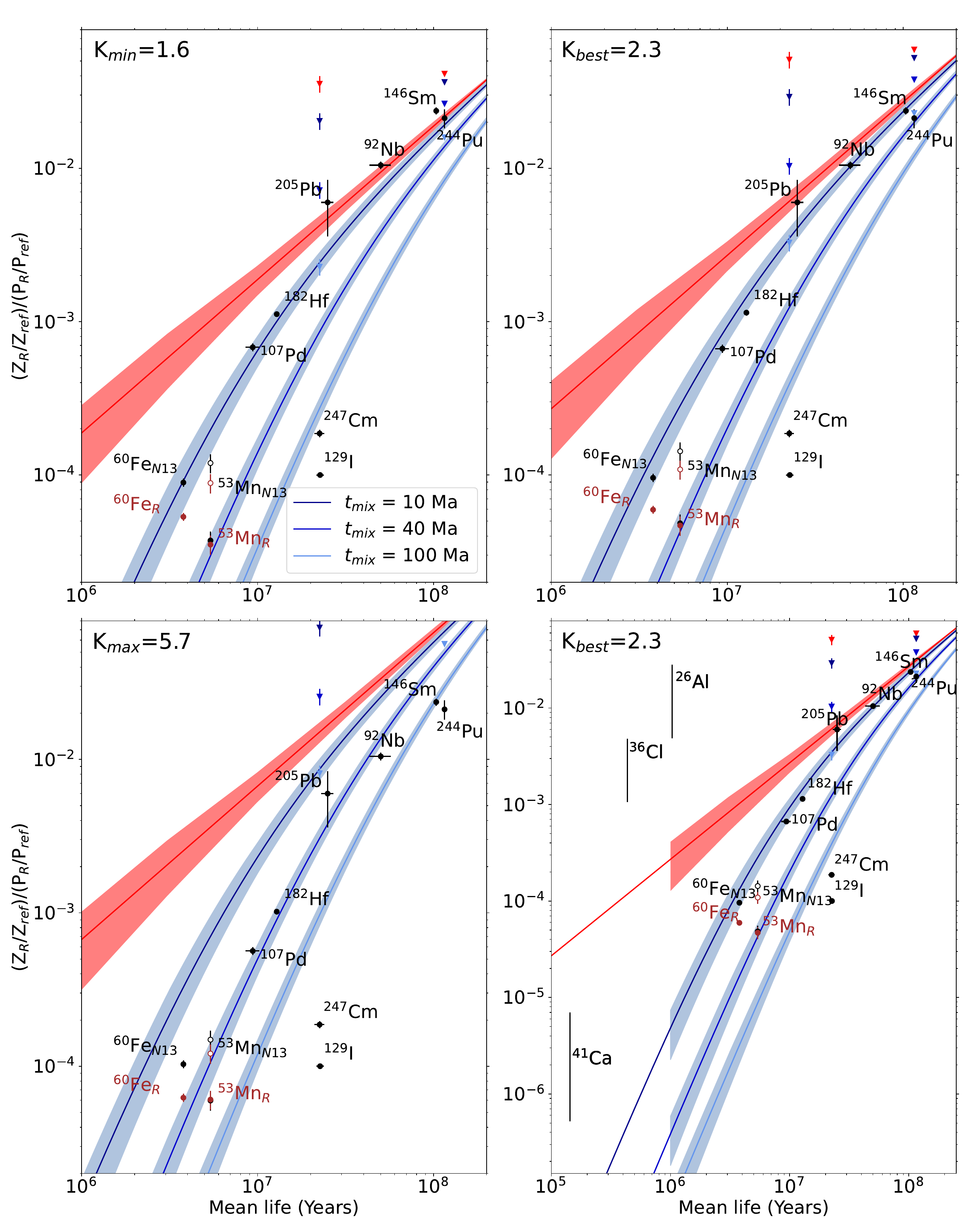}
\caption{Same as Fig.~\ref{fig:fig1} (and Fig.~\ref{fig:fig1zoomout} in the bottom right panel), but with the blue lines representing the predictions from the mixing scenario of \citet{Clayton1983}, described in Sec.~\ref{sec:methodsteady} for different values of the mixing timescale, $t_{\mathrm{mix}}$.}
\label{fig:fig3}
\end{figure*}

In all the figures (Fig.~\ref{fig:fig1}, Fig.~\ref{fig:fig1zoomout}, and Fig.~\ref{fig:fig3}), we plot the ESS abundance ratios over their corresponding production ratios ($(Z_R/Z_{\mathrm{ref}})/(P_R/P_{\mathrm{ref}})$) as a function of the mean life. The predicted ISM values of the steady-state scenario (at a Galactic age of 8.5 Ga) and their Monte-Carlo uncertainties (see Sec.~\ref{sec:methodsteady}) are marked with red lines and surrounding shaded regions. As \iso{247}Cm and \iso{244}Pu require a different steady-state equation (see Sec.\ref{sec:methodsteady}), their steady-state values are represented by downward-facing triangles. Additionally, all figures show the measured ESS abundance ratios over their corresponding production ratios (described in Sec.~\ref{sec:methodinput}). Black circles mark these measured values. In Fig.~\ref{fig:fig1}, we also show the results of the free-decay model (see Sec.~\ref{sec:timescales}) marked by blue lines (and blue downward-facing triangles for \iso{247}Cm and \iso{244}Pu). These lines correspond to those predicted for the ESS and should therefore be compared directly to the points. Additionally, in Fig.~\ref{fig:fig1} (and Fig.~\ref{fig:fig3}), three subplots corresponding to different values of $K$ are shown, as indicated on top of each panel. Figure~\ref{fig:fig1zoomout} reproduces the middle panel of Fig.~\ref{fig:fig1} but with the x-axis extended down to $\tau$=0.1 Ma to include the very short-lived SLRs. Figure~\ref{fig:fig3} is the same as Fig.~\ref{fig:fig1}, except that the mixing timescale is implemented instead of the free-decay model.  

As mentioned in the Introduction, these figures are useful to examine consistency among the SLRs; in other words, to determine if any of the models presented in Sec.~\ref{sec:timescales} can self-consistently explain the ESS abundances of the SLRs. A self-consistent solution means that all the abundances can be matched using the same $t_{\mathrm{iso}}$ or $t_{\mathrm{mix}}$, for a given, fixed value of $K$, given that different $K$ values represent a different realisation of the Milky Way. Practically, this means that all the black data points -- representing the measured meteoritic data -- should be located within error bars on the same blue line -- representing a given model.

Figure~\ref{fig:fig1} shows that reasonable, self-consistent solution can be found for \iso{53}Mn, \iso{60}Fe, \iso{107}Pd, \iso{182}Hf, and \iso{205}Pb, when considering $K$=1.6 and 2.3. For the solution to work for \iso{53}Mn, it is required to use the GCE result, where the yield of this isotope from Type Ia SNe is set to zero. Figure~\ref{fig:fig1} shows that the solution is close to a free-decay time of 10 Ma, and we checked that indeed the best solutions for $K=1.6$ and 2.3 are found with free-decay times of 9 and 12 Ma, respectively.

For $K=1.6$, the \iso{53}Mn and \iso{60}Fe CCSN yields from N13 are slightly preferred, as the R case does not provide as good a solution for \iso{53}Mn. Instead, for $K=2.3$, the CCSN yields from R are slightly preferred, as the N13 case does not provide as good a solution for \iso{60}Fe. The \iso{53}Mn result is affected by the fraction of Type Ia SNe assumed to be of sub-Chandrasekhar type, and choosing a lower fraction than the 0.5 value used here would result in slightly lower \iso{53}Mn/\iso{55}Mn production ratios, likely closer to the fitting lines.

The results of our analysis are in agreement with those shown in Figures 4 and Extended Data Figure 4 of \citet{leckenby24} and Figure 5 of \citet{trueman2025}, 
which plot the probability distribution of the combined error bars we reported on the y-axis (both the shaded regions and the errors on the dots) as a function of the free-decay time. The fact that these five isotopes are reasonably explained using the GCE alone indicates that there is no need for a local source to the presolar molecular cloud, unless the free-decay time was instead much longer (>50 Ma) and the GCE signature was erased. However, such a long timescale would be at odds with the lifetimes of molecular clouds, of up to 30 Ma. Moreover, CCSN models polluting the ESS do not provide a self-consistent solution and generally overproduce the isotopes considered here \citep{vescovi2018, battino2024}. 

For the $r$-process nuclei, it is impossible to find a self-consistent solution since \iso{129}I, \iso{244}Pu, and \iso{247}Cm are consistently below their predicted values. This confirms once again that these SLRs originated from a last event \citep{Cote_2021}.

The $p$-process \iso{92}Nb and \iso{146}Sm are in agreement with each other only in the $K=2.3$ case, where they also agree with the other SLRs. Instead, in the $K=1.6$ case, they both require a non-physical, negative free-decay time. For the $K=5.7$ case, they disagree both with each other and with the other SLRs. Note that these results depend on the half-life of \iso{146}Sm: for instance, using the value of 92 Ma \citep{chiera2024}, no solution is found between the two SLRs. This issue, as well as the stellar origin of these nuclei, still needs to be investigated\footnote{We checked the possible location of the $p$-process nuclei \iso{97}Tc and \iso{98}Tc in Fig.~\ref{fig:fig1} using the production ratio derived in the same way as for the other $p$-process nuclei. Because their ESS values are only upper limits, the location of these isotopes would be represented by a downward arrow in the figure (the same applies to \iso{135}Cs), which we found to not provide further constraints.}.

As shown in Fig.~\ref{fig:fig1zoomout}, the very short-lived SLRs \iso{26}Al, \iso{36}Cl, and \iso{41}Ca are significantly more abundant in the ESS than predicted by the steady-state model, indicating that this approach cannot account for their presence. Note that for \iso{41}Ca, even a free decay of only 1 Ma would not allow a match. 
This confirms previous results from GCE modelling \citep{Huss2009, sahijpal2014evolution, Cote2019a} and indicates the need for a local source of these isotopes within the pre-solar molecular cloud. The winds of massive stars are a favourite scenario \citep{gaidos2009,ku2022}, because, as mentioned above, a CCSN does not appear to produce a satisfactory fit. The latest results from the single and binary models of \citet{brinkman2021,brinkman2023} show that current predictions of the composition of the winds from single and binary stars of mass above roughly 35 \msun\ can reproduce the ESS \iso{26}Al/\iso{27}Al and \iso{41}Ca/\iso{40}Ca ratios, while the \iso{36}Cl/\iso{35}Cl ratio appears to require a minimum higher initial stellar mass. We note that these stars can also produce \iso{107}Pd and \iso{205}Pb \citep{arnould2006}, and this requires further investigation.  

Figure~\ref{fig:fig3} shows the prediction from the mixing model (see Sec.~\ref{sec:timescales}), to be compared to the free-decay model shown in Fig.~\ref{fig:fig1} and Fig.~\ref{fig:fig1zoomout}. Figure~\ref{fig:fig3} shows that reasonable, self-consistent solutions -- i.e., solutions with the same mixing time, within the respective error bars -- can be found for \iso{53}Mn, \iso{60}Fe, \iso{107}Pd, \iso{182}Hf, and \iso{205}Pb. Figure~\ref{fig:fig3} also shows that the solution is close to a mixing time of 10 Ma, and we checked that indeed the best solutions for $K$=1.6 and 2.3 are found with mixing times of 11 and 13 Ma, respectively. The $K=1.6$ solution does not include \iso{205}Pb. For \iso{53}Mn and \iso{60}Fe, in the $K=2.3$ case, the N13 CCSN yields are slightly preferred, as the R yields cannot provide a satisfactory solution. Different from the free-decay time scenario, it is possible to find a reasonable self-consistent solution also in the case of $K=5.7$, with a much longer mixing time of 38 Ma, and a preference for the R yields. As in the case of Fig.~\ref{fig:fig1}, the GCE scenario in which the yield of \iso{53}Mn from Type Ia SNe is set to zero is the only way to obtain a match between \iso{53}Mn and the other isotopes.

According to the formulas of \citet{Clayton1983}, the timescale derived here ($t_{\mathrm{mix}}$) corresponds to the time by which the total mass of each ISM phase would be fully replenished, depending on the total mass and the mass exchange rate between the different phases (see their Eq. 1). In fact, Clayton's model assumes that the molecular cloud would constantly be replenished and therefore it would never disperse. However, now we know that giant molecular clouds have a limited lifetime between roughly 10 and 30 Ma due to disruption by stellar feedback \citep[see review by][]{chevance2020}, although they could also accrete material and live longer lives \citep{jeffreson24}. Therefore, while Clayton's model needs to be updated, the values for the mixing timescale we found here seem reasonable because they are comparable to the observed lifetimes of giant molecular clouds. Combining the two pieces of information (i.e., that the cloud material would have been completely replenished within the value of the $t_{\mathrm{mix}}$ and that giant molecular clouds are known to have lifetimes within such range) we infer that the Sun was born in a cloud that lived up to roughly 10 Ma in the $K=1.6$ and 2.3-type Milky Way Galaxies or roughly 40 Ma in a $K=5.7$-type Galaxy. 

The $r$-process isotopes are again well below the predicted values. Only \iso{244}Pu can potentially be explained, but the mixing time would be unrealistically high compared to the other isotopes (we remind the reader that for this isotope, the downward triangles need to be checked rather than the lines). Similar to the case of the free-decay scenario, here the $p$-process \iso{92}Nb and \iso{146}Sm are in agreement with each other and the rest of the SLRs only in the $K=2.3$ case. In the cases of the shortest-lived SLRs, the picture is not significantly improved from Fig.~\ref{fig:fig1zoomout}.

Based on the discussion above, both scenarios -- when coupled with the SLR data -- constrain the lower mass of the stars that evolved and exploded within the cloud before the Sun formed. From the three galaxies and two models, we have six distinct configurations, five of which point to maximum lifetimes of roughly 9--12 Ma, corresponding to the evolutionary timescale of stars of masses >15--20 \msun\ and one case points to maximum lifetimes of roughly 40 Ma, corresponding to the evolutionary timescale of stars of masses >8 \msun. These values may be used to constrain the origin of the very short-lived SLRs (e.g., \iso{26}Al) from stellar sources that could evolve within the cloud’s lifetime. Recent observational studies of molecular cloud lifetimes, however \cite{chevance2020} show that massive stars tend to form late during the cloud’s evolution, thereby limiting the available feedback window to at most $\sim$5 Myr. This would increase to roughly 35 \msun\ the minimum mass of the local stellar source of the very short-lived SLRs \citep[in agreement with][see discussion of Fig.~\ref{fig:fig1zoomout}]{brinkman2023}. The other potential solution involves no stellar feedback within the cloud but a strongly enhanced contribution from massive star winds within the steady-state scenario \citep{yang2012, desch2023}. This scenario will need to be carefully evaluated, considering the contribution to the ISM of realistic stellar yields from single and binary stellar models, including rotation, and the most likely values of the free-decay times and mixing time scales derived here from the other SLRs.

\section{Conclusions}
\label{conclusion}

We have presented an updated version of a commonly used visual representation to compare predicted and observed abundance ratios of SLRs relative to their respective stable (or long-lived) isotopes, in which ratios are normalised to their production ratios and plotted against their mean life. We assumed that the SLRs were in a steady-state equilibrium and considered two scenarios to predict the ESS abundances. In the first approach, the pre-solar molecular cloud inherited SLRs from the ISM, but it was separated from it. During the time of this separation, the SLRs underwent exponential decay until the formation of the first solids in the ESS. In the second approach, the different phases of the ISM experienced some mixing, leading to a quadratic decrease of SLR abundances within the pre-solar molecular cloud. The timescale factors -- the isolation time and the mixing time -- are free parameters. We compared the predictions of the two models to ESS abundances derived from meteoritic analysis. For this comparison, we used the most up-to-date input for the ESS abundances, the mean life, and stellar production ratios (see Sec.~\ref{table:big}). 

We found that: 
\begin{enumerate}
    \item[1)] The ESS abundances of the isotopes produced by the $s$-process (\iso{107}Pd, \iso{182}Hf and \iso{205}Pb) can be accounted for by inheritance from the ISM enriched only by AGB stars. In this case, the required isolation time is around 10 Ma if the $K$ parameter has the minimum or best value, while no viable solution is found if $K$ has the maximum value. The required mixing time is around 10 Ma for the minimum and best value of $K$, while it increases to approximately 40 Ma if $K$ has the maximum value.
    \item[2)] The ESS abundances of the explosive SLR isotopes (\iso{53}Mn and \iso{60}Fe) show a similar pattern to the $s$-process isotopes, i.e., they require a similar isolation or mixing time. We used two sets of production ratios for these SLRs, and did not find strong constraints on them as both sets produced acceptable results depending on the other parameters (such as the value of $K$ and the fraction of sub-Chandrasekhar Type Ia SNe). We found that the \iso{53}Mn isotope can only be matched consistently with all the other isotopes if its yield from Type Ia SNe is set to zero, which means that the last Type Ia SN occurred at least a few tens of Ma before the formation of the ESS \citep[see also][]{trueman2025}.
    \item[3)] The ESS abundances of the $p$-process isotopes (\iso{92}Nb and \iso{146}Sm), when assumed that they originate from Type Ia SNe only, are occasionally consistent with the $s$-process and the explosive SLR isotopes, but only when $K=2.3$ and the half-life of \iso{146}Sm is taken to be around the lowest limit of the range of the currently available values. 
    \item[4)] The ESS abundances of the $r$-process isotopes (\iso{129}I, \iso{244}Pu, and \iso{247}Cm) would require much longer timescales, in agreement with their origin from a last event instead of steady-state ISM \citep{Cote_2021}.
    \item[5)] The ESS abundances of the very short-lived SLRs (\iso{26}Al, \iso{36}Cl and \iso{41}Ca) are much higher than any predictions based on steady-state ISM, which supports the conclusions of previous studies that a different explanation is required for their origin.
\end{enumerate}

Several factors limit the conclusions of this paper or impede the making of further conclusions. These factors include uncertainties in the ESS abundances and the mean lives, although these are unlikely to alter the overall picture displayed here significantly. The most pressing question concerns the origin of the $p$-process isotopes; we only included their production from Type Ia SNe, even though they are also known to be produced in CCSNe \citep{roberti23gamma, Roberti_2024}. 
Moreover, better ESS data for the $p$-nuclei \iso{97}Tc and \iso{98}Tc would help establish a more coherent view for $p$-nuclei. Future work should also consider the nuclear physics uncertainties, which could alter stellar production ratios, such as the rate of the \iso{22}Ne($\alpha$,n)\iso{25}Mg reaction, which affects the production of \iso{182}Hf. 

We have focused on the steady-state scenario, which is valid if $\tau/\delta \gtrsim 2$, therefore, implying that the $\delta$ for CCSNe needs to be quite short ($\sim1$ Ma). This contrasts with the prediction of \citet{Wehmeyer_2023}, who suggested a value of $\sim$10 Ma. In this case, scenarios other than the steady state should be statistically investigated for \iso{53}Mn and \iso{60}Fe. In particular, single or multiple, but discrete, injections into the ISM parcel of gas that became the pre-solar cloud need to be evaluated \citep[see][for discussion on the last $s$-process event]{Trueman_2022}. While the timing of a last event does not provide information on the isolation time nor mixing time, it is still useful as an upper limit for them. 

Finally, if the very short-lived isotopes originated from self-polluting events within the molecular cloud (rather than the ISM), our results indicate that the most likely candidate stellar sources lived less than roughly 10 Ma, implying a mass above 20 \msun. It is necessary to test such candidates against their possible contribution to the other isotopes considered here.

\begin{acknowledgments}

We thank Uli Ott for the discussion and suggestions, and the anonymous referee for their careful reading and the many comments, which helped to substantially improve the manuscript.
This work was supported by the the Lend\"ulet Program LP2023-10 of the Hungarian Academy of Sciences and the European Union’s Horizon 2020 research and innovation programme (ChETEC-INFRA -- Project no. 101008324) and the NKFI via K-project 138031 (Hungary). B.S. thanks the research assistant program of Konkoly Observatory. M.L. was also supported by the NKFIH excellence grant TKP2021-NKTA-64. T.T. acknowledges support from the ERC Synergy Grant Programme (Geoastronomy,
grant agreement number 101166936, Germany).

\end{acknowledgments}


\end{document}